\documentclass[aps,prb,reprint,groupedaddress,floatfix]{revtex4-1}

\usepackage[english]{babel} 
\usepackage{amssymb,amsmath}
\usepackage{graphicx}
\usepackage{url}
\usepackage{multirow}
\usepackage{chemformula}
\usepackage{tabularx}
\usepackage{makecell}
\usepackage{xcolor}
\usepackage{siunitx}

\usepackage{subcaption}

\newlength{\smallpic}
\begin{document}

\title{Data-driven discovery of novel 2D materials by deep generative models}
\author{Peder Lyngby$^*$ and Kristian Sommer Thygesen
 }
 \address{Computational Atomic-scale Materials Design (CAMD), Department of Physics, Technical University of Denmark, 2800 Kgs. Lyngby Denmark \\ \normalfont{$^*$Corresponding author: pmely@dtu.dk}}
\date{\today}

\begin{abstract}
 Efficient algorithms to generate candidate crystal structures with good stability properties can play a key role in data-driven materials discovery. Here we show that a crystal diffusion variational autoencoder (CDVAE) is capable of generating two-dimensional (2D) materials of high chemical and structural diversity and formation energies mirroring the training structures. Specifically, we train the CDVAE on 2615 2D materials with energy above the convex hull $\Delta H_{\mathrm{hull}}<\SI{0.3}{eV/atom}$, and generate 5003 materials that we relax using density functional theory (DFT). We also generate 14192 new crystals by systematic element substitution of the training structures. We find that the generative model and lattice decoration approach are complementary and yield materials with similar stability properties but very different crystal structures and chemical compositions. In total we find 11630 predicted new 2D materials, where 8599 of these have $\Delta H_{\mathrm{hull}}<\SI{0.3}{eV/atom}$ as the seed structures, while 2004 are within 50 meV of the convex hull and could potentially be synthesized. The relaxed atomic structures of all the materials are available in the open Computational 2D Materials Database (C2DB). Our work establishes the CDVAE as an efficient and reliable crystal generation machine, and significantly expands the space of 2D materials. 
\end{abstract}

\maketitle

\section{Introduction}
The discovery of new materials that meet specific requirements e.g. in terms stability, compatibility, or physical properties, is an exciting scientific challenge of great relevance for our society.  First-principles quantum mechanical calculations, e.g. based on density functional theory (DFT)\cite{kohn1965self}, can predict the structure and properties of materials with high accuracy even before they are made in the lab. However, a DFT code by itself is insufficient for realising the paradigm of inverse materials design.

Considering the vast number of possible materials and the complexity of general structure-property relations, it becomes clear that successful inverse design relies on the following critical components: (i) automated execution and management of large numbers of atomistic calculations, (ii) access to large amounts of relevant high quality materials data, and (iii) efficient algorithms that can propose new candidate materials from data. In addition, synthesis and characterisation experiments must be included in the loop as well, but this aspect will not be considered here.    

Components (i) and (ii) are largely in place. Indeed, the advent of workflow management engines for computational materials science\cite{curtarolo2012aflow,jain2015fireworks,pizzi2016aiida,gjerding2021atomic,mortensen2020myqueue} have made it possible to perform high-throughput (HT) computations for thousands of materials with minimal human intervention\cite{greeley2006computational,madsen2006automated,curtarolo2013high,kirklin2013high,ornso2013computational,zhang2019computational,chen2016understanding,hachmann2011harvard,bhattacharya2015high,castelli2012computational,hautier2013identification,yu2012identification,kuhar2018high,aykol2016high,mounet2018two,chen2015design}. Atomic structures and basic materials properties from such HT studies have been stored in computational databases\cite{thygesen2016making,himanen2019data,saal2013materials,jain2013commentary,borysov2017organic,curtarolo2012aflow,draxl2019nomad,haastrup2018computational,cheon2017data,gjerding2021recent}, which together contain results of millions of DFT calculations. Complemented by experimental crystal structure databases, this makes a rich and rapidly growing data source for materials science. 

The main challenge concerns component (iii). In previous HT studies, the candidate materials to be explored were mostly produced by lattice decoration of known reference materials. An obvious limitation of this approach is that the resulting materials by construction will be similar to the reference materials. In particular, the 3-tuple: (space group, occupied Wyckoff positions, stoichiometry) is invariant under element substitution. 

Generative machine learning algorithms could potentially broaden the diversity of candidate materials beyond the lattice decoration paradigm. 
However, designing a successful generative model for periodic materials has proved challenging due the problem of creating representations of the lattice, atomic coordinates and elemental composition that are both invariant to translations and rotations and is invertible\cite{Noh2020Machine}. The vast chemical space of elements that can be present in inorganic crystals further complicates the design of representations. Therefore, previous implementations of generative models for periodic materials have either been limited to a fixed subset of chemical elements \cite{Noh2019Inverse,Kim2020Generative,Long2021Constrained} and/or a subset of possible crystal structures \cite{Zhao2021High,Song2021Computational}. Recently, a general invertible representation has been proposed\cite{Ren2022invertible}, which encodes the material as a matrix of both real and reciprocal space features, but is not invariant under translations and rotations. Xie \textit{et al.} developed a crystal diffusion variational autoencoder (CDVAE) model \cite{xie2021crystal}, which uses a generative diffusion model to circumvent the need for an invertible representation and employs an equivariant graph neural network to ensure invariance (in fact, equivariance).

\begin{figure*}
    \centering
    \includegraphics[scale=0.5]{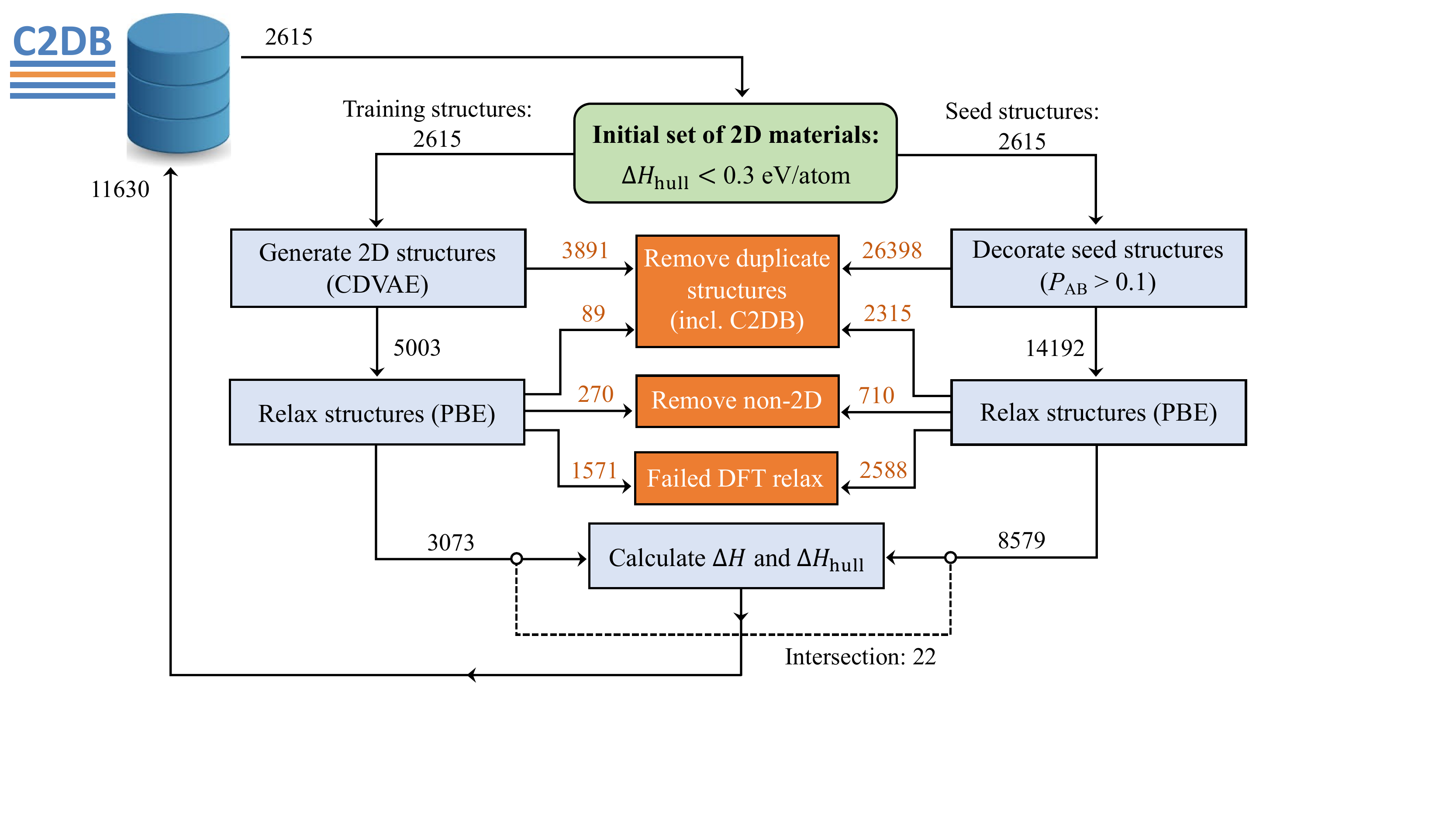}
    \caption{Workflow to generate candidate 2D materials using the CDVAE generative model (left branch) and lattice decoration (right branch). The same set of 2615 materials is used to train the CDVAE model and as seed structures for lattice decoration, respectively. Black numbers indicate the number of materials present at a given step of the workflow while orange numbers indicate number of materials discarded.}
    \label{fig:workflow}
\end{figure*}

In this work, we train a CDVAE\cite{xie2021crystal} on 2615 2D materials with formation energy up to 0.3 eV/atom from the convex hull, and generate 10000 two-dimensional (2D) crystals. We compare these structures to a set of 14192 2D crystals obtained by systematic lattice decoration of the training structures. While Ref. \cite{xie2021crystal} assessed validity and diversity of the generated crystals by means of qualitative measures, such as charge neutrality and minimum bond distance, we here conduct a systematic, unbiased quantitative analysis by performing full DFT-based relaxations and stability analysis of the generated structures. Compared to the crystals in the training set, the structures generated by the CDVAE (after DFT relaxation) show similar formation energies but significant differences in both composition and crystal structure. In general, CDVAE seems able to produce more complex materials without compromising stability. 

As a direct test of the CDVAE model's capacity to learn the stability properties of the training structures, we also train a CDVAE on materials lying at least 0.4 eV/atom above the hull. We find that the structures generated by this model have significantly higher formation energies that those produced by the CDVAE trained on the more stable materials. 

In addition to providing a quantitative assessment of the CDVAE, our work identifies no less than 8599 new unique 2D materials with an energy above the convex hull below $\SI{0.3}{eV/atom}$ many of which could potentially be synthesised. The generated crystal structures are freely available as part of the C2DB\cite{haastrup2018computational}.

\section{Results and discussion}

\begin{figure*}
\centering
\includegraphics[width=1\linewidth]{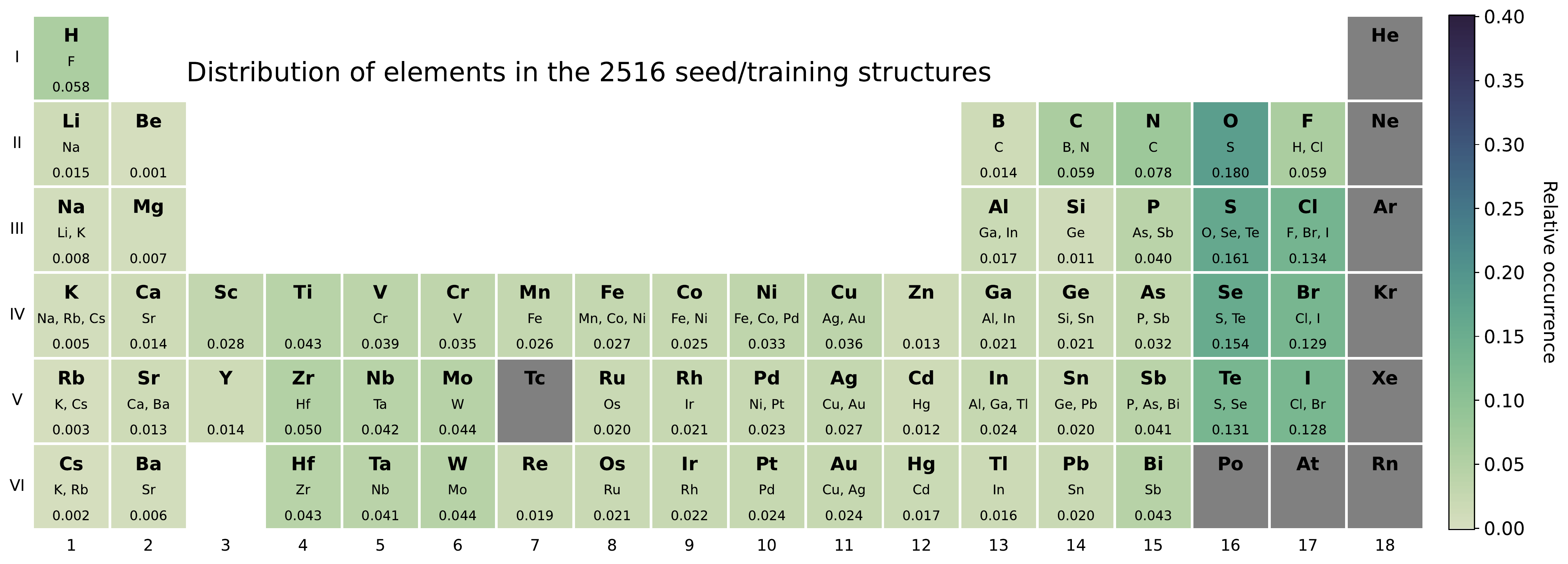}
\caption{Heat map of the relative occurrence of each element in the 2D materials used to train (seed) the CDVAE (LDP). The middle row shows the element substitutions for the LDP corresponding to $P_{AB}>0.1$. The relative occurrence is shown in the last row.}
\label{fig:pt}
\end{figure*}

\subsection{Crystal diffusion variational autoencoder}

The CDVAE combines a variational autoencoder\cite{kingma2013autoencoder} and a diffusion model to generate new periodic materials. The crystal is represented by a tuple consisting of the atomic number of the $N$ atoms, their respective coordinates, and the unit cell basis vectors. CDVAE consist of three networks: the encoder, a property predictor, and the decoder. The encoder is a SE(3) equivariant periodic graph neural network (PGNN), which encodes the material onto a lower dimensional latent space from which the property predictor predicts the number of atoms $N$, the lattice vectors, and the composition, which is the fraction present of each element.  The decoder is a noise conditional score network diffusion model\cite{song2019generative} that takes a structure with noise added to the atom types and coordinates and learns to denoise it into the original stable structure. Here the score is an estimate of the gradient of the underlying probability distribution of the materials and is predicted by another SE(3) equivariant PGNN. The three networks are trained concurrently.

New materials can be generated after training by using the property predictor to sample the latent space. A unit cell with the predicted basis vectors is then initialised with the predicted atoms placed at random positions. Using the decoder, the atom types and coordinates of the initial random unit cell are then gradually denoised into a material that is similar to the data distribution of the training data. CDVAE utilizes that adding noise to a stable material will likely decrease its stability and, thus, by learning to denoise the noisy stable structure, the decoder learns to increase the stability of the structure. Therefore CDVAE should be trained only on stable materials.
An in-depth description of CDVAE can be found in Xie \textit{et al.} \cite{xie2021crystal}.

The set of materials used as training data for the CDVAE and seed structures for the lattice decoration protocol (LDP), respectively, consists of 2615 unique 2D materials from the C2DB\cite{haastrup2018computational,gjerding2021recent}. As our aim is to discover new stable materials we limited the initial set of materials to the subset of C2DB with energy above the convex hull $\Delta H_{\mathrm{hull}}< \SI{0.3}{eV/atom}$. This was done because both the CDVAE (LDP) are more likely to generate stable materials when trained on (seeded by) stable materials. We did not exclude dynamically unstable materials.

After training the CDVAE model, 10.000 structures were generated of which 1106 failed CDVAE's basic validity check (charge neutrality and bond lengths above 0.5 Å). Of the remaining 8894 structures, 3891 are duplicate structures which are sorted out (see section \ref{sec: Method} for more details) and the rest are relaxed using DFT.

\subsection{Lattice decoration protocol}

The lattice decoration protocol (LDP) substitutes the atoms in the seed structures by atoms of similar chemical nature. As a measure of chemical similarity we use the probability matrix $P_{AB}$ introduced by Glawe \textit{et al.}\cite{glawe2016optimal}, which describes the likelihood that a stable material containing a chemical element $A$ remains stable after the substitution $A\to B$. Glawe \textit{et al.} constructed this probability matrix based on an analysis of materials in the Inorganic Crystal Structure Database\cite{belsky2002new}. We choose a substitution probability of 10 \% ($P_{AB}>0.1$), which generates the substitutions shown in Fig. \ref{fig:pt}. Based on these substitution relations, we perform all possible single and double substitutions for all seed structures. For example, the seed structure MoS$_2$ generates six MX$_2$ structures with M=Mo,W and X=O, S, Se (the seed structure itself included). The total set of resulting materials are analysed for structures that share the same reduced formula and space group. Such structures are considered as duplicate structures and are filtered out. After removal of duplicates, we are left with 14192 unique 2D crystals (the seed structures excluded) which are relaxed using DFT.

\begin{figure*}
\centering
\includegraphics[width=1\linewidth]{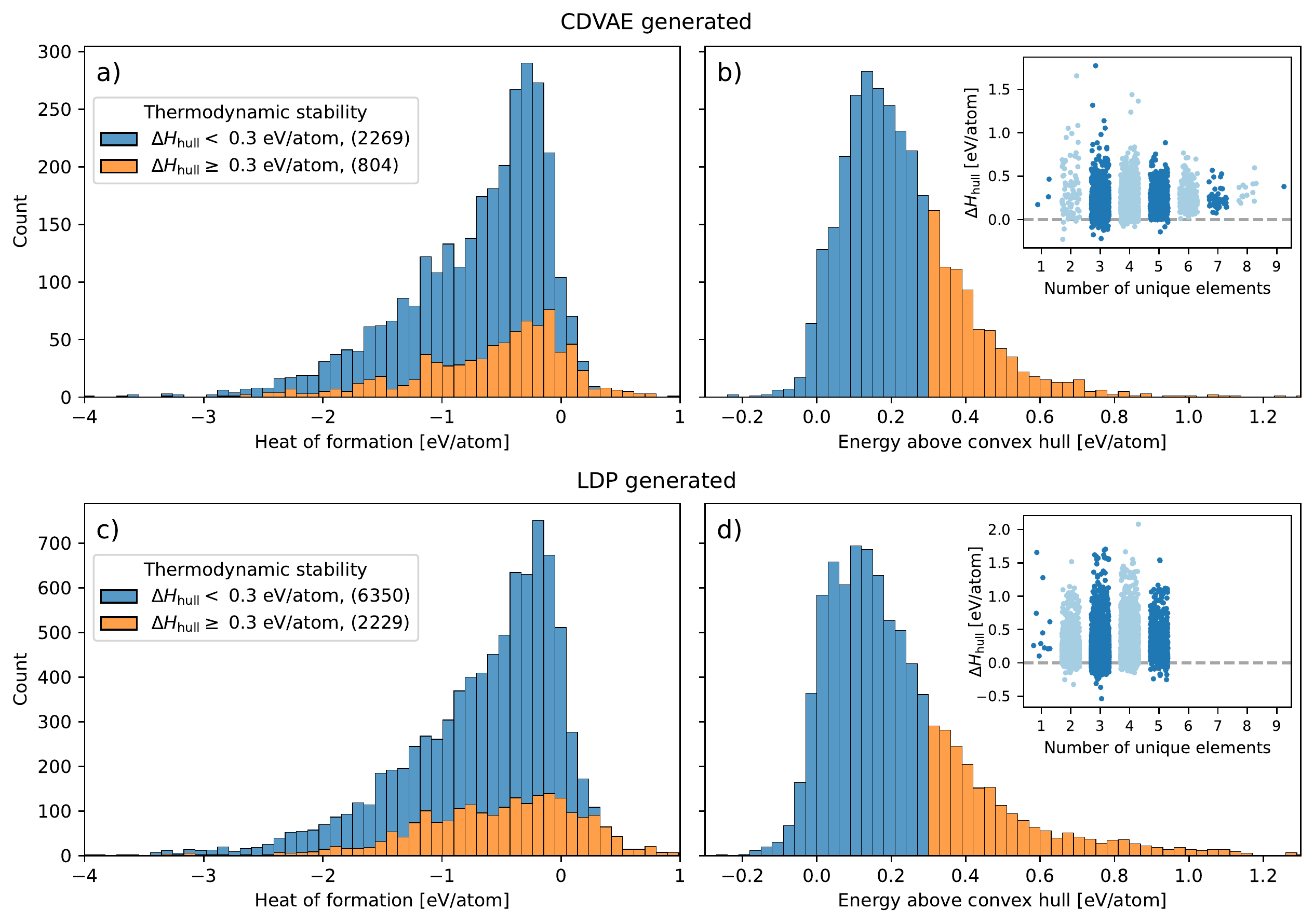}
\caption{Histogram of the heat of formation and energy above convex hull for the DFT-relaxed structures resulting from the CDVAE (a, b) and LDP (c, d) methods. The inset shows the energy above convex hull with respect to the number of unique elements in the structure.}
\label{fig:hist_thermo}
\end{figure*}

\subsection{Workflow}
Our workflow is illustrated in Fig. \ref{fig:workflow}. Starting with the initial set of 2D materials, we generate two new sets of crystal structures using CDVAE and LDP, respectively. Duplicate structures within each set are removed (see section \ref{sec: Method} for more details). The now unique crystal structures are relaxed using DFT calculations employing the PBE xc-functional (see section \ref{sec: Method} for more details). After the relaxation, any new duplicate structures are removed again and as are materials that have relaxed into non 2D structures (we refer to Ref. \cite{gjerding2021recent} for details on the dimensionality analysis). Finally the heat of formation, $\Delta H$, and the energy above convex hull, $\Delta H_{\mathrm{hull}}$, are calculated.

\begin{table}[]
\begin{tabularx}{0.45 \textwidth}{ 
   >{\raggedright\arraybackslash\hsize=.6\hsize}X 
   >{\centering\arraybackslash\hsize=.2\hsize}X 
   >{\raggedleft\arraybackslash\hsize=.2\hsize}X  }
\hline
                     & LDP   & CDVAE                          \\ \hline
Success rate         & 82 \%                       & 69 \%                        \\
Avg. number of steps & 40.1       & 55.5\\
Avg. energy decrease $\left[  \si{eV/atom} \right]$ & 0.62  & 0.51  \\ \hline
\end{tabularx}
\caption{Summary statistics for the DFT relaxation of the two methods for generating initial structures.}
    \label{tab:statistics}
\end{table}

\begin{figure}[t]
    \centering
    \includegraphics[width=1\linewidth]{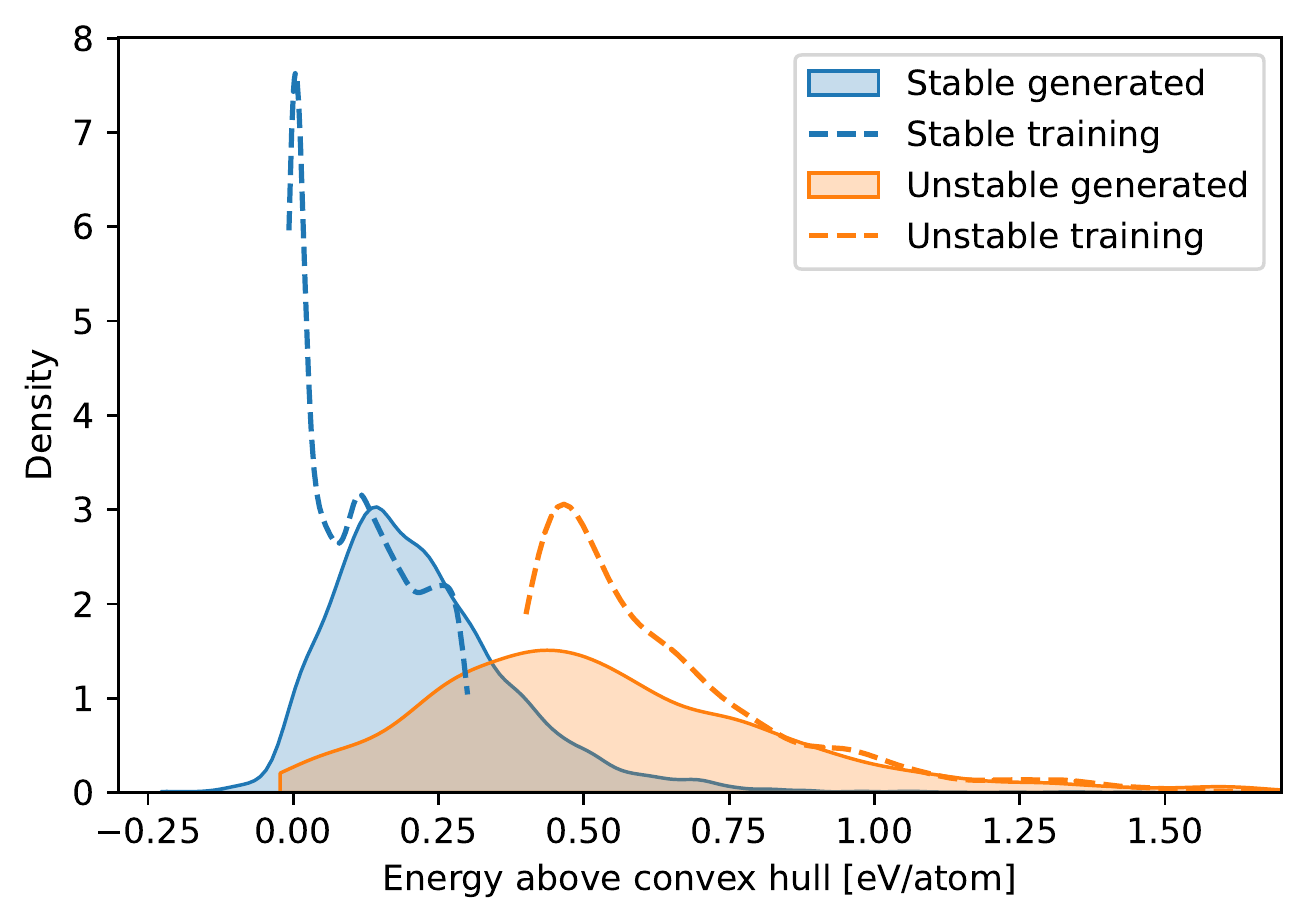}
    \caption{Kernel density estimate showing the distribution of the convex hull energies for the stable and unstable CDVAE generated dataset as well as their training data.}
    \label{fig:kde}
\end{figure}

\begin{figure*}
\centering
\includegraphics[width=1\linewidth]{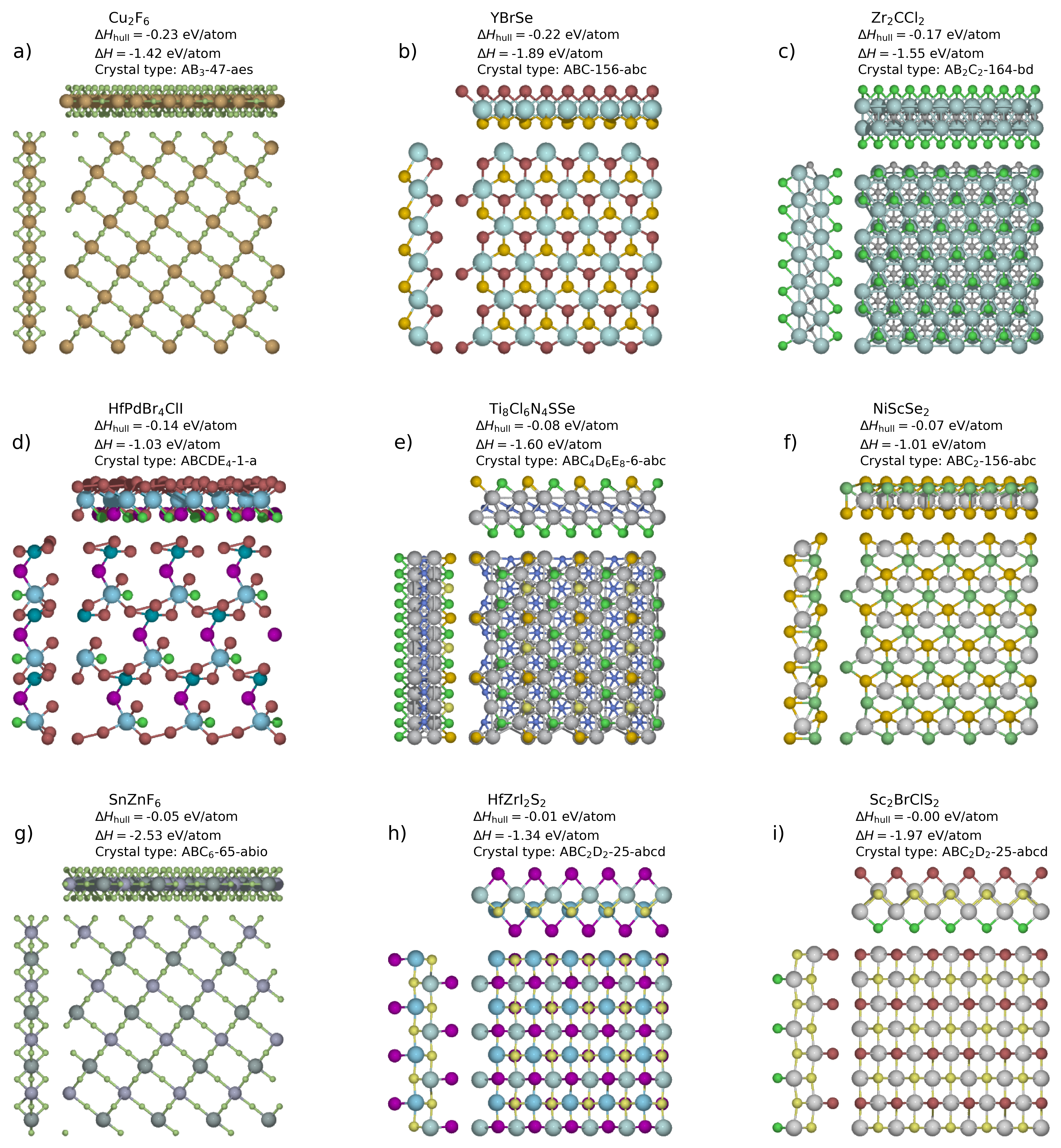}
\caption{(a-g) Examples of CDVAE generated materials with negative convex hull energies. (h, i) Examples of CDVAE generated stable materials with the new discovered combination of stoichiometry ABC$_2$D$_2$, space group number 25 and occupied Wyckoff positions a,b,c,d.}
\label{fig:cdvae_examples}
\end{figure*}

In Table \ref{tab:statistics} we report the success rates for the DFT relaxations of the structures generated by CDVAE and LDP, respectively, together with the average number of relaxation steps and the average energy decrease from the initial to the relaxed structure. All three parameters are assumed to describe 
how close the initial structures are to the final DFT relaxed structures - e.g. a structure from a perfect generative method would only need one relaxation step and the energy decrease would be zero. As expected, neither LDP or CDVAE generate stable relaxed structures. However, while the LDP on average requires less steps to relax, the CDVAE structures are closer in energy to the relaxed structure. The fact that the number of relaxation steps and reduction in energy upon relaxation is comparable for LDP and CDVAE, suggest that the CDVAE-generated crystals are as close to relaxed structures as the LPD-generated structures. 

We observe that the DFT relaxation fails for about 18 \% of the LDP-generated and about 31 \% of the CDVAE-generated structures. The vast majority of these failures are due to problems in converging the Kohn-Sham SCF cycle. We suspect that a large fraction of the convergence problems occur for materials with magnetic ground state (all calculations are performed with spin polarisation). This is supported by the fact that 30 \% of the materials containing one or more of the magnetic 3d-metals (V, Cr, Mn, Fe, Co, Ni), fails due to convergence errors, while this is only happens for 10 \% of other materials.

\subsection{Thermodynamic stability}

A histogram of the heat of formation and the energy above the convex hull for the (DFT-relaxed) structures resulting from the CDVAE and LDP are shown in Fig. \ref{fig:hist_thermo}. The distributions of both $\Delta H$ and $\Delta H_{\mathrm{hull}}$ obtained for the two structure generation methods are remarkably similar. For example, 73.8 \% of the CDVAE materials have $\Delta H_{\text{hull}}$ below $\SI{0.3}{eV/atom}$ (as the training data) while this is the case for 74.0 \% of the LDP materials. It should, however, be noted that the smaller success rate of the DFT relaxation of the CDVAE generated materials could influence these statistics as it likely that many of the structures which could not be converged would have resulted in unstable structures. The inset of Fig. \ref{fig:hist_thermo} shows how the energy above the convex hull is distributed depending on the number of different elements in the structure. First of all it is evident that CDVAE is able to create structures with a larger number of unique elements than is present in the training data (5 unique elements is the maximum in the seed structures), while LDP is limited to the stoichiometries present in the seed materials. However, generally the thermodynamic stability is lower for the materials with larger number of unique elements. Examples of some of the most stable CDVAE generated structures is shown in Fig. \ref{fig:cdvae_examples}. The material Zr$_2$CCl$_2$ shown in c) is one of the 22 materials which are found both by the CDVAE and LDP method.

To predict whether a given 2D material can be synthesized is a complex problem that involves many factors. Often the size of $\Delta H_{\mathrm{hull}}$ is used a soft criterion for synthesizability as it determines the material's thermodynamic stability relative to other competing phases (this criterion neglects growth kinetics and substrate interactions both of which can be important for 2D materials). A previous study of 700 polymorphs in 41 common inorganic bulk material systems showed that a threshold of $\Delta H_{\mathrm{hull}}<0.1$ eV/atom will exclude 26\% of the known synthesized polymorphs\cite{aykol2018thermodynamic}. We also note that the T-phase of MoS$_2$ was synthesised both as a monolayer\cite{kappera2014phase} and a layered bulk\cite{bell1957preparation}, despite having $\Delta H_{\mathrm{hull}}=0.18$ eV/atom\cite{urlc2db}. These examples demonstrate that many of the predicted 2D materials with $\Delta H_{\mathrm{hull}}<50$ meV/atom (2004) or even $\Delta H_{\mathrm{hull}}<100$ meV/atom (3400), are likely to be synthesizable.  

While the $\Delta H_{\mathrm{hull}}$-distributions in Fig. \ref{fig:hist_thermo} are clearly peaked close to zero they also have a tail of less stable materials. In particular, about 26 \% of the materials have $\Delta H_{\text{hull}}>\SI{0.3}{eV/atom}$ (the threshold to select the training structures). A natural question to ask is then to what extent the structures produced by the CDVAE are in fact biased towards high stability structures? To answer this question, we trained a CDVAE model on 988 2D materials with a $\Delta H_{\text{hull}}>\SI{0.4}{eV/atom}$ and used it to generate another 10.000 structures from which we randomly selected 1000 non-duplicate structures, which we relaxed following the same workflow as described before. The distribution of the energy above the convex hull of the relaxed structures for both the stable and unstable CDVAE models are shown in Fig. \ref{fig:kde} together with the distribution of their respective training sets. We clearly see that the CDVAE model trained to generate unstable materials produces structures that are significantly further from the convex hull than the stable model. This illustrates that CDVAE successfully learns the chemistry of the materials in the training data.

\begin{figure*}
\centering
\includegraphics[width=1\linewidth]{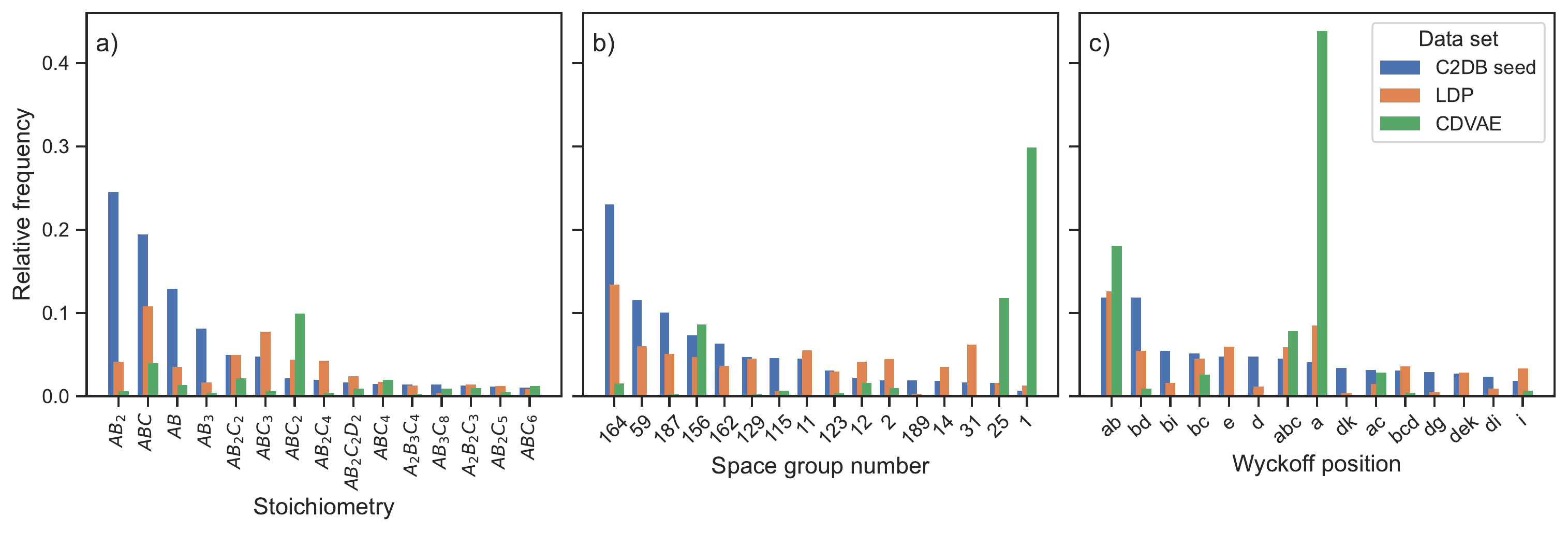}
\caption{Relative frequency of the stoichiometry, space group number and occupied Wyckoff positions for each of the data set.}
\label{fig:structure}
\end{figure*}

\subsection{Structural diversity}

Having established the capability of the CDVAE to produce materials with good stability properties, we now turn to its ability to generate crystals of high chemical and structural diversity. While the LDP is restricted to stoichiometries and crystal structures already present in the seed structures, the CDVAE (in principle) has no such limitations. Fig. \ref{fig:pt} shows the relative occurrence of each element in the seed/training structures. The corresponding plots for the materials generated by LDP and CDVAE (after relaxation) are shown in Fig. 1 in the Supplemental Information. Both LDP and CDVAE produces diverse compositions with elements covering most of the periodic table. However, CDVAE has a significantly higher occurrence of oxygen and chalcogens (S and Se) as well as halogens (Cl, Br and I). This trend is also present for the materials prior to relaxation and, thus does not originate from a potential higher DFT convergence rate for these elements. Instead, the six elements are also more prevalent, albeit slightly, in the seed structures which could indicate an overfitting of the model.

\begin{figure}
\centering
\includegraphics[width=1\linewidth]{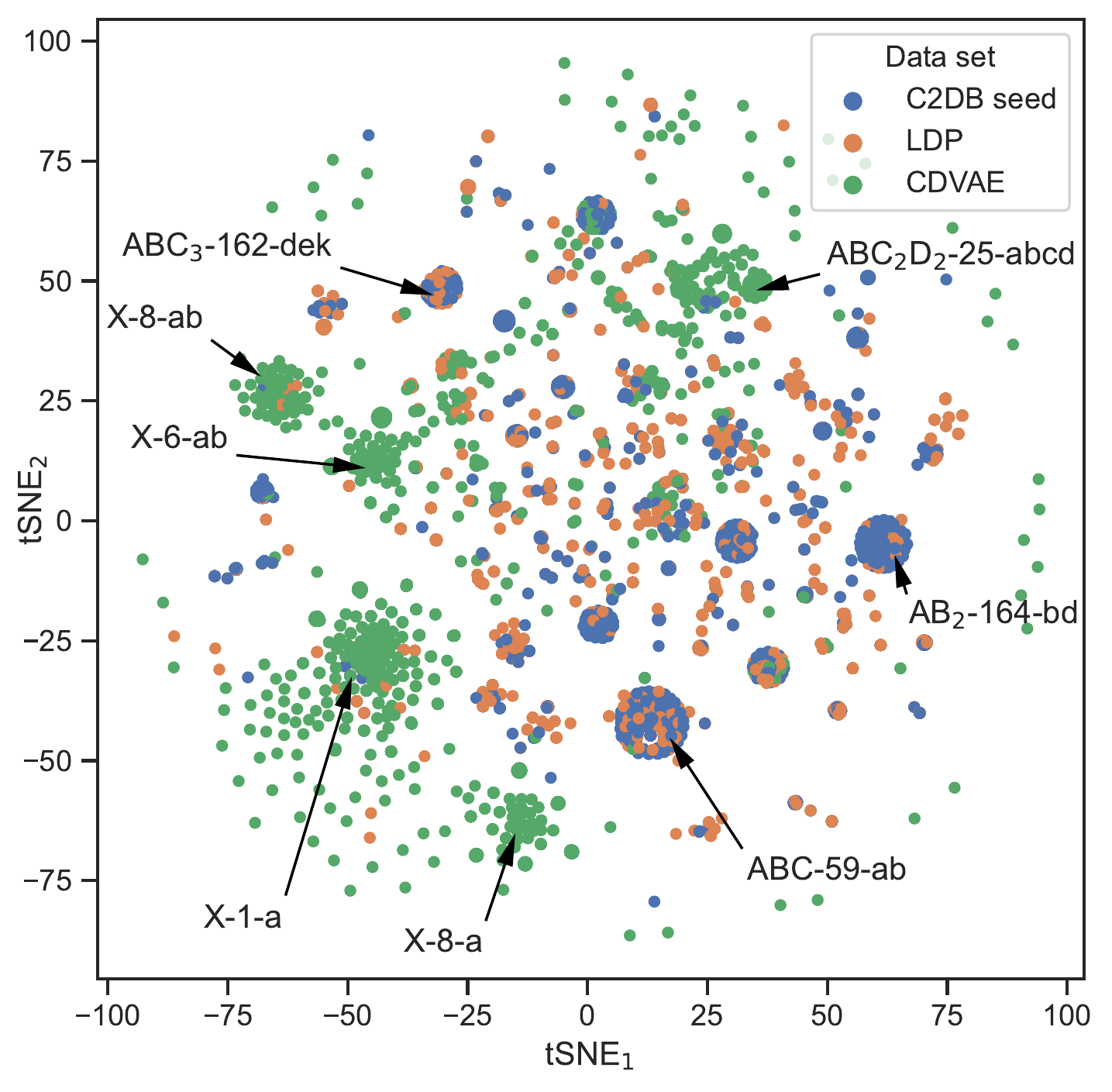}
\caption{t-SNE embedding of the structure
represented as a individual one-hot encoding of the stoichiometry, space group, and occupied Wyckoff positions. Selected clusters are highlighted as 'stoichiometry'-'space group'-'Wyckoff position'. 'X' corresponds to an  arbitrary stoichiometry.}
\label{fig:tSNE}
\end{figure}

The CDVAE generates significantly different chemical compositions and crystal structures as compared to the seed structures and those generated by the LDP. Fig. \ref{fig:structure} shows the relative frequencies of stoichiometry, space group number and occupied Wyckoff positions, respectively. Only the most common classes of the seed structures are shown. We find 239 unique stoichiometries among the CDVAE-generated materials, while there is only 87 and 103 unique stoichiometries in the seed structures and LDP-generated structures, respectively. The higher number of unique stoichiometries in the LDP-generated structures than in the seed structures is due to new stoichiometries being created when two different elements are substituted by the same element, or when an element is being substituted with an element already present in the seed material. For example, the seed materials Te$_2$Cu$_4$O$_{12}$ (stoichiometry AB$_2$C$_6$) becomes Cu$_4$S$_{14}$ (stoichiometry A$_2$B$_7$) under the double substitution O$\to$S and Te$\to$S.
The significantly larger number of unique stoichiometries generated by CDVAE compared to the LDP shows that the former is able to produce new classes of structures that are not present in the training data.

The CDVAE tends to generate rather complex, low-symmetry structures, which is illustrated by the large fraction of materials with space group number 1. Moreover, the average number of different elements in the unit cell is 4.0 for the CDVAE generated materials while it is only 2.6 for the C2DB seed structures. The larger number of different elements is part of the reason for the higher fraction of materials with low symmetry. This tendency of CDVAE to generate structures with more complex composition is also noted by Xie \textit{et al.}, who attributes this to a non-Gaussian distribution of the underlying structure of the materials. Thus, when CDVAE generates new materials it samples from a Gaussian distribution $\mathcal{N}(0,1)$ from which it predicts the number of atoms and composition. However if $\mathcal{N}(0,1)$ is not representative of the latent space, out of distribution materials can be generated. For materials discovery this could, however, be advantageous as this makes CDVAE able to generate new crystal types which are not present in the training data.

To give a global overview of the structural distribution of the three data sets, a t-SNE embedding is shown in Fig. \ref{fig:tSNE}. The t-SNE analysis is made for 2500 materials sampled randomly from each data set. Here the structure is represented as a tuple given by the space group, occupied Wyckoff positions, and stoichiometry, where each is one-hot encoded before the t-SNE embedding. We see that most of the training data form clear clusters, which represent the most common stoichiometries, space group and Wyckoff positions. The LDP generated materials mostly follow the same pattern as the seed structures. However, the CDVAE generated structures are more spread out, which is partly due to the large variation in their stoichiometries, while a few clusters appear due to the large fraction of low symmetry materials with space group number 1. One noteworthy example is the cluster of CDVAE generated materials with stoichiometry ABC$_2$D$_2$, space group number 25 and occupied Wyckoff positions a,b,c,d. For this specific combination, CDVAE discovered 123 new materials of which 30 lies within 50 meV of the convex hull, while there is no examples of such materials in the training set nor in the LDP generated structures. Two of the most stable discovered materials of this type can be seen in Fig. \ref{fig:cdvae_examples} (h, i). The new class of structures have broken out-of-plane symmetry either due to the outermost atoms (b) or the innermost atoms (a). The fact that the CDVAE is able to generate new classes of stable materials, which are not present in the training data, is very promising and a clear advantage of deep generative models compared to lattice decoration protocols.

\section{Conclusions}
In conclusion, we have successfully employed a deep generative model in combination with a systematic lattice decoration protocol (LDP) to generate more than 8500 unique 2D crystals with formation energies ($\Delta H$) within 0.3 eV/atom of the convex hull. Out of these, more than 2000 have $\Delta H$ within 50 meV/atom of the convex hull, and could potentially be synthesized. This represents at least a doubling of the known stable 2D materials.  

In addition to the very significant expansion of the known space of 2D materials, our work provides a quantitative assessment of the crystal diffusion variational autoencoder (CDVAE)\cite{xie2021crystal}, and establishes its excellent performance with respect to the two key criteria: ability to learn the stability properties of the training structures, and ability to generate crystals with high chemical and structural diversity. In fact, only 25\% of the generated materials had $\Delta H_{\mathrm{hull}}$ above the 0.3 eV/atom threshold used to select the training structures, and the stoichiometries of the generated materials span 239 types versus 87 present in the training structures. Generally, the crystal structures generated by CDVAE have higher complexity and lower symmetries than the training structures. We found the method of lattice decoration to be complementary to the CDVAE generator with the two methods yielding only 22 common crystals out of the 11630 structures generated in total. While the LDP is limited to the structural blueprint of the seed materials, CDVAE is able to generate new classes of materials, which are not present in the initial data set. This is promising for an autonomous materials discovery method as it adds new genes to pool of trial materials and thus goes beyond the lattice decoration paradigm. 

The fact that CDVAE is comparable to lattice decoration (with substitution by chemically similar elements) in terms of stability while producing new and diverse crystal structures, is a testimony to the prospect of using deep generative models in materials discovery. 

All the structures are available in the C2DB database\cite{urlc2db}, and their basic properties will also be made available as the execution of the C2DB property workflow proceeds.

\section{Method} \label{sec: Method}

\subsection{Workflow}
To set up and manage the workflow we use the Atomic Simulation Recipes\cite{gjerding2021atomic}, which has implemented tools for DFT relaxation, duplicate removal, dimensionality check, and for calculating the thermodynamic properties. The DFT calculations are performed using the GPAW code \cite{Mortensen2005real} with the PBE xc-functional, a plane wave cut-off energy of $ \SI{800}{eV}$ and a \textit{k}-point density of at least  $\SI{4}{Å}$. The relaxation is stopped when the maximum force is below $ \SI{0.01}{eV/Å}$ and the maximum stress is below $ \SI{0.002}{eV/Å^3}$.

The duplicate removal recipe finds duplicate structures using the root mean square distance (RMSD) between the structures which is calculated using the Python library pymatgen\cite{ong2013Python}. We consider structures to be duplicate if RMSD$<0.3$ Å and only keep the structure with the lowest heat of formation. See Ref. \cite{gjerding2021recent} for more information. For the initial LDP generated materials (before the DFT relaxation) a more crude duplicate sorting of the structures is employed, where two materials with the same reduced formula and space group are considered duplicates.\\

To determine the convex hull we use as reference databases the C2DB as well as a database of reference structures comprising 9590 elementary, binary, and ternary crystals that all lie within 20 meV of the convex hull in the Open Quantum Materials Database (OQMD)\cite{saal2013materials}. These reference structures were relaxed using the VASP\cite{kresse1993ab} code at the PBE level (PBE+U for selected transition metal oxides) as part of the OQMD project. Since we use the GPAW code to relax and evaluate the energy of the 2D materials, we have re-calculated the total energy of the reference structures (without re-optimisation) using the GPAW code. 

\subsection{CDVAE}
CDVAE is designed to generate 3D bulk crystals, where the unit cell is periodic in all three directions. This introduces a problem when generating 2D materials, which are non-periodic in one direction. We solve this issue by introducing an artificial periodicity in the non-periodic direction with a lattice vector which is an order of magnitude larger than those in the periodic directions. This ensures that the graph networks only connect atoms in the 2D layer and thus CDVAE learns to generate 2D materials. When training the model, we used 70 \% of the materials in the training set, while 15 \% was used for validation and 15 \% for test. We used the same hyperparameters as employed by Xie \textit{et al.} for their MP-20 data set. See Ref. \cite{xie2021crystal} for more information.

\section{Acknowledgements}
We thank Morten N. Gjerding for assistance with setting up the lattice decoration protocol. 
We acknowledge funding from the European Research Council (ERC) under the European Union’s Horizon 2020 research and innovation program Grant No. 773122 (LIMA) and Grant agreement No. 951786 (NOMAD CoE). K. S. T. is a Villum Investigator supported by VILLUM FONDEN (grant no. 37789).

\section{Author contributions}
P.L. and K.S.T. developed the initial concept. P.L. ran the generative models, the DFT simulations and performed the data analysis. K.S.T. supervised the project and aided with the interpretation of the results. P.L. and K.S.T wrote and discussed the paper together.

\section{Competing interests}
The authors declare no competing interests.

\section{Data availability}
All the discovered crystal structures and their properties will be available as a part of C2DB (\url{https://cmr.fysik.dtu.dk/c2db/c2db.html})

\bibliography{refs}

\end{document}